# Gravitational Starlight Deflection Measurements during the 21 August 2017 Total Solar Eclipse

**Donald G. Bruns**
7387 Celata Lane, San Diego, CA 92129

E-mail: dbruns@stellarproducts.com

## Abstract

Precise starlight positions near the Sun were measured during the 21 August 2017 total solar eclipse in order to measure their gravitational deflections. The equipment, procedures, and analysis are described in detail. A portable refractor, a CCD camera, and a computerized mount were set up in Wyoming. Detailed calibrations were necessary to improve accuracy and precision. Nighttime measurements taken just before the eclipse provided cubic optical distortion corrections. Calibrations based on star field images 7.4° on both sides of the Sun taken during totality gave linear and quadratic plate constants. A total of 45 images of the sky surrounding the Sun were acquired during the middle part of totality, with an integrated exposure of 22 seconds. The deflection analysis depended on accurate star positions from the USNO's UCAC5 star catalog. The final result was a deflection coefficient L = 1.752 arcsec, compared to the theoretical value of L = 1.751 arcsec, with an uncertainty of only 3%.

## 1. Introduction

In 1919, Sir Arthur Eddington attempted to measure the starlight deflection caused by the Sun's gravity [Dyson et. al. (1920)]. He was trying to test Albert Einstein's recent calculations based on his General Theory of Relativity, taking advantage of a total eclipse of the Sun to image stars. The deflections should appear radially from the center of the Sun and decrease as the reciprocal of that distance, with a coefficient L = 1.751 arcsec for a star at the Sun's limb. After detailed examinations of several large photographic plates, Eddington concluded that Einstein was right. His pronouncement made Einstein a world-wide celebrity and a household name. Over the years that followed, however, and after similar measurements taken at later eclipses, the measurement uncertainty in the deflections was never reported better than about 6% [von Klüber (1960)]. Some experiments have been re-analyzed in attempts to remove perceived biases in favor of Einstein, making those results much more uncertain. Not until 1970's-era radio frequency measurements and recent astrometric satellite measurements were Einstein's deflections precisely measured [Will (2015)]. Figure 1 shows the results of the initial reports for every eclipse experiment, as well as the newer radio frequency measurements.

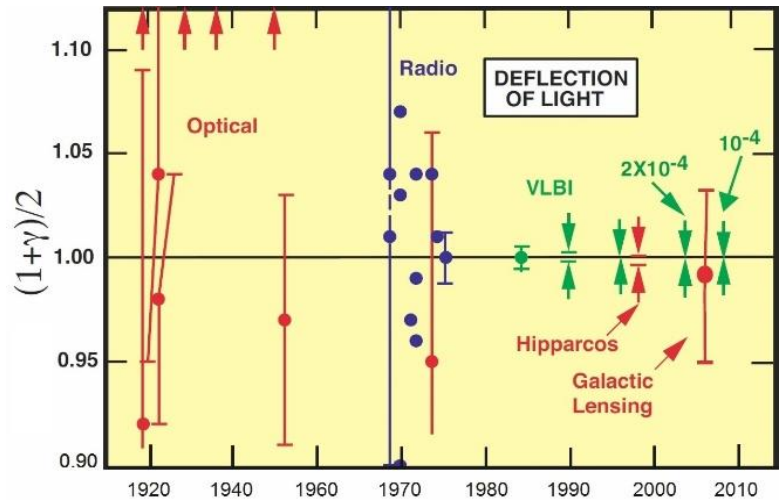

**Figure 1.** The relative value of the deflection coefficient and its uncertainty are shown for optical (red), radio (blue), and VLBI (green) experiments. In general relativity, the space curvature constant γ equals one. All previous eclipse experiments took place before 1973 and never resulted in better than apparent 6% precision. Four experiments resulted in values that lie above the top of this graph. The best optical measurement come from the Hipparcos satellite where the uncertainty was 0.2%. Gaia satellite data is expected to reach 0.0001%. Figure is adapted from Will (2015).

The measurable stars closest to the Sun have deflections typically 0.3 arcsec to 1 arcsec. In order to measure these with even 10% precision, the locations of the stars need to be determined to within 0.03 arcsec to 0.10 arcsec. These small numbers are what makes the experiment so difficult. From 1919 until the last successful measurement in 1973 [Jones (1976) and Brune et. al. (1976)], optical measurements during seven total solar eclipses used the best available technology [von Klüber (1960)]. This meant large photographic plates (0.2 m to 0.45 m), long refractor telescopes (1.5 m to 8.5 m focal length), and comparison images taken with the same telescope a few months before or after the eclipse to determine non-perturbed star positions. One of their key desires was to take plate scale calibration images during totality to remove one of the largest error sources. Astronomers attempted several innovative techniques, including the use of beamsplitters to simultaneously image these calibration fields. The experiments were always very difficult, resulting in technical failures. Weather caused some of the problems, of course, but the 20th century astronomers always hoped that someday the experiment would be completed without errors, providing much more accurate results [Freundlich and Ledermann (1944)].

The recent 21 August 2017 total solar eclipse across the United States provided a convenient opportunity to repeat this experiment. This paper reports successful starlight deflection measurements performed in Wyoming using high quality amateur astronomical equipment. The final results show the most precise and accurate measurements of this kind ever reported. The calculated uncertainty is only 3.4%, suggesting the deflection coefficient L is likely to be between 1.69 arcsec and 1.81 arcsec. The final coefficient was measured to be 1.752 arcsec, which is within 0.05% of the theoretical value.

One of the key improvements that simplified this experiment was the availability of absolute star positions with 0.002 arcsec precisions. This eliminated star measurements of comparison images taken months away from the eclipse, thus eliminating some error-prone data analysis steps. Corrections for atmospheric refraction were still needed, but those were easily calculated from local weather measurements. Hence, this particular trial had fewer obstacles to reach high accuracy.

Using current technology generated two new problems. Atmospheric turbulence apparent in short exposures causes star images to wander randomly between frames. The magnitude of this effect depends on the atmospheric seeing, exposure length, and telescope aperture [Lindegren (1980), Zacharias (1996)]. Each star's motion is uncorrelated, since they are separated by relatively large angles from each other. The apparent motion measured 0.4 arcseconds RMS for the 0.62 second exposures used during totality, comparable to their gravitational deflections. All previous eclipse experiments used photographic plates and large focal ratios that eliminated this problem because the star wander was integrated over their long exposures. This broadened the star images, but did not seriously affect their positions. Modern CCD sensors necessitate nominal exposures less than 1 second to prevent pixel saturation, so all of the images were first aligned and then averaged together to simulate a single long exposure. This technique reduced the centroid measurement error.

The other new problem was getting correct exposures. With photographic plates, the final image could be monitored in the darkroom during chemical development. The technician stopped the process before too much darkening ruined the image. CCD sensors have a much smaller effective dynamic range. Short exposures make star centroid positions uncertain due to digital noise while overexposure saturates some pixels, preventing accurate measurements. Fortunately, current portable computers are fast enough to run an autoexposure program with only a small timing penalty. While analysis of images from previous eclipses was used to estimate the ideal exposures and led to predictions that made this experiment seem feasible, implementing autoexposure contributed to its successful completion.

## 2. Methods

### 2.1 Equipment

A review of commercially available telescopes and cameras led to selections of the Tele Vue Optics, Inc. NP101is refractor and the Finger Lakes Instrumentation, Inc. ML8051 CCD camera. A Software Bisque MyT Paramount provided celestial tracking. The technical criteria are detailed in this section. Table 1 summarizes the equipment parameters and Figure 2 shows the observation location in Wyoming.

A refractor avoids a central obscuration that adds scattered light and reduces contrast. The stellar image needs to be about two pixels diameter; a larger image diameter reduces pixel contrast and so might not reveal enough stars while a smaller image diameter might make determining centroid locations impossible. The preferred refractor was a Petzval design so that an internal optical alignment scheme could be implemented [Bruns (2017)]. These requirements narrowed the telescope choices down to the Tele Vue NP101is and the Takahashi FSQ106. The optical distortion of the NP101is was known to be small, based on the author's previous experience with this telescope, so that telescope was selected for this experiment. To improve the critical off-axis performance, the outer 7 mm of the front aperture was masked, making an effective F/6.2 instrument with an 87 mm aperture.

**Table 1.** The equipment chosen for this experiment, along with parameter values and some explanatory details.

| Equipment parameter | Parameter value | Parameter details |
|---|---|---|
| Telescope model | NP101is | Tele Vue Optics, Inc. |
| Telescope design | Nagler-Petzval refractor | Front and rear doublet lenses |
| Telescope aperture | 101 mm | Masked to 87 mm |
| Telescope focal length | 543 mm | Design value (at 622 nm) |
| Camera model | ML8051 | Finger Lakes Instrumentation, Inc. |
| Camera sensor | KAI-08051 | Interline CCD with lenslets |
| Sensor operating temp. | -20 C | TE cooled |
| Camera digitizer | 16 bits | 0 to 65535 ADU counts |
| Camera digitizing rate | 12 MHz | 0.7 sec to digitize image |
| Camera overhead time | 1.43 s | Full-frame USB2 download/save |
| Sensor pixel array | 3296 (H) x 2472 (V) | Single digitizer output |
| Sensor pixel dimensions | 5.5 µm | Lenslet array increases efficiency |
| Pixel Field of View | 2.087 arcsec | Calculated from images |
| Sensor Field of View | 1.9° by 1.4° | Right Ascension is along rows |
| Mount model | MyT Paramount | Software Bisque, Inc. |
| Mount controller | TheSky X | Includes T-Point mount model |

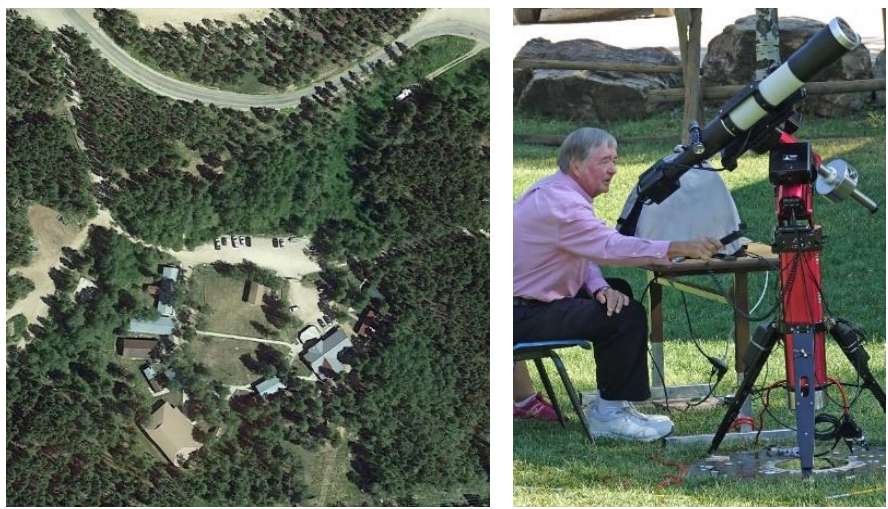

**Figure 2.** (Left) The setup location was near the top of Casper Mountain. The Lions Camp consists of an array of small buildings surrounded by trees. The telescope was located in the grassy field near the center of the figure. Image is from Google Earth. (Right) The NP101is refractor and ML8051 CCD camera are supported by a MyT Paramount on its field tripod bolted to a base fixed in concrete. The scale is shown by comparison with the author. Image courtesy Steve Lang.

Comparing different CCD alternatives led to the use of a monochrome interline sensor. This guaranteed identical exposures on all pixels and avoided potential shutter vibration effects when used with short exposures. The desired field of view needed to be large enough to contain nearby stars out to about five solar radii, or 1.3° from the Sun. Using a larger array would have added significant overhead time by imaging stars that had very small deflections, probably contributing only noise to the solution. A monochrome camera allowed more accurate centroids. The camera needed to be cooled to reduce thermal noise. The only sensors that met these requirements were the ON Semiconductor KAI-08051 and the Sony ICX694 series CCD used by several vendors in their astronomical cameras. The author had previous experience with a Finger Lakes Instrumentation KAI-series camera, so that model was chosen. A 16-bit readout is used in the camera, providing output signals from 0 to 65535 ADU (analog-to-digital unit) counts. Only one digitizer connects to the sensor, so the entire array is read with the same gain.

To reduce chromatic aberration from the telescope as well as reduce atmospheric dispersion differences between different spectral–class stars, an r' Sloan filter from Astrodon was bonded in place just in front of the camera sensor. This filter also increased the contrast between the blue-tinted sky and the redder stars which made up most of the measureable targets. Reducing the amount of light into the sensor slightly increased the exposure time, but this actually reduced atmospheric turbulence effects. Most of the camera time is spent on digitizing, downloading, and saving the images, so adding the color filter did not significantly affect the timing. Figure 3 shows the sensor response curve and the red filter transmission.

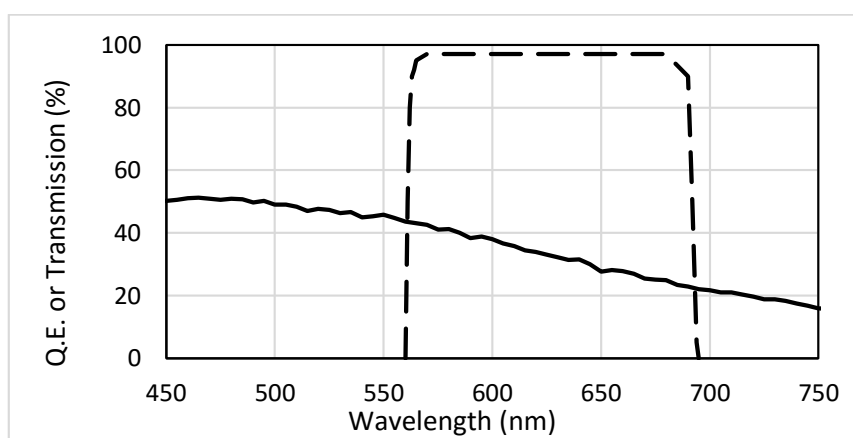

**Figure 3.** The ON Semiconductor KAI-08051 interline CCD response curve is shown between 450 nm and 750 nm, slowly dropping from 50% to 16%. The dashed curve is the transmission of the r' Sloan filter. When these curves are convoluted with the typical red spectrum of the dimmer target stars, the equivalent central wavelength was 622 nm.

The telescope mount needed to be a portable equatorial design, so the author used a Software Bisque, Inc. MyT Paramount. This computerized mount easily handled the mass of the chosen telescope and camera and was easily set up on its matching tripod in Wyoming. Its sub-arcsecond periodic error and sub-arcminute all-sky pointing accuracy assured that good images would be acquired. Fortunately, a good polar alignment (according to T-Point mount modelling software, to about four arcminutes) was completed the Friday night before the Monday morning eclipse. This reduced image rotation between the calibration fields down to a few arcseconds. The field tripod legs were securely bolted to a 0.9 m diameter eclipse-commemorating concrete base, ensuring a setup stable against wind and other moving hazards.

The experiment location was chosen based on the desire for a high altitude and the probability of clear weather. The high altitude was intended to reduce atmospheric turbulence and provide bluer skies. The Allen H. Stewart Lions Camp on Casper Mountain in Wyoming is located at an elevation of 2390 m and has adjoining lodging facilities. The specific setup location was surrounded by a grassy field and tall trees, which also improved local seeing and provided a mitigation against wind. During a visit in August 2016, a solar scintillometer from AiryLab [Seykora (1992), Beckers (1992)] was used to estimate the daytime turbulence. While this measurement was only relative and did not provide the seeing parameter $r_0$, the values at this location were three times better than locations measured a day later at lower altitudes in Nebraska. Figure 2 shows the location of the setup 13 km south of Casper.

The weather conditions at the site were monitored so that atmospheric refraction could be corrected in the final calculations [Stone (1996)]. Two small digital thermometers were set up, one with a one-second response time in air and one with a one-minute response time in air. The barometric pressure was monitored with a local instrument and verified by comparison with the nearby Casper airport. Refraction is not very sensitive to the dew point, but this was measured on Casper Mountain within 2 C, using a calibrated hygrometer.

## 2.2 Software

Commercial software packages were used for many of the experimental and analysis tasks in this experiment. Astrometrica and Diffraction Limited's MaxIm DL were used to semi-automatically or manually determine the dozens of centroids in the eclipse images. More details are explained in the Section 2.3. All of the calculations were kept separate between the two programs. Since there was no *a priori* reason to give a higher confidence to one program over the other, it was decided before the analysis was completed to report the weighted average of the individual program results.

Optical distortion corrections required analyzing a long series of pre-eclipse nighttime images. Because these optical distortion image series contained thousands of stars, MaxIm DL was not used. Instead, Astrometrica and DC-3's Visual PinPoint were successfully used. Both of these programs automatically find centroids using different curve-fitting routines. The average of the Astrometrica- and PinPoint-determined distortion coefficients were used for the MaxIm DL centroid analysis. The effect of averaging those coefficients is estimated to change the reported deflection result by less than 0.001 arcsec.

Software Bisque's TheSkyX ran the telescope mount and MaxIm DL operated the camera. A Visual Basic script sent commands to adjust the exposures, timing, and the mount pointing direction. An autoexposure routine controlled the exposure durations and timing of all of the image series. For the data analysis, a combination of FORTRAN and MathSoft's Mathcad were used along with Microsoft Excel to keep track of the parameters and intermediate results.

## 2.3 Centroid Calculations

Since the stellar image diameters (measured at the full-width at half-maximum intensity [FWHM]) are nominally 1.5 pixels or 3.1 arcsec, the location of each stellar centroid must be determined to within a small fraction of one pixel to get the desired precision. Two analysis software programs, Astrometrica and MaxIm DL, were compared to determine how accurately they can determine the centroids. Astrometrica determines centroids by fitting a Gaussian curve to the pixel values across a radial center while MaxIm DL is based on a moment calculation.

Typical stars are shown in Figure 4, enlarged to show individual pixels. The star in the left image has an SNR near the minimum used in this experiment, while the one on the right has an SNR three times larger. The left star is nearly centered on a pixel while the right star straddles the border between two pixels. Both stars are brightness scaled for this figure so that the peak pixel is white. The black level is scaled to show the noise level near the star. The inner ring, three pixels in radius, is the boundary used by MaxIm DL to determine which pixels (or fraction) to use in its moment technique. The annulus between the rings is used to determine the background level. The same two stars were analyzed in the Astrometrica program. That program shows the radial curve fit and below it, the residual error. A two-pixel radius was selected in Astrometrica to perform its curve fitting. The difference in these centroids between the two programs in the horizontal direction was only 0.039 pixels and 0.003 pixels, while the differences in the vertical direction were 0.017 and 0.002 pixels. The difference was small enough that there was no statistical reason to choose one program over the other. The mean RMS residual error for all of the stars in a 10 s long image in the nighttime optical distortion image series was typically 0.03 pixels for both programs.

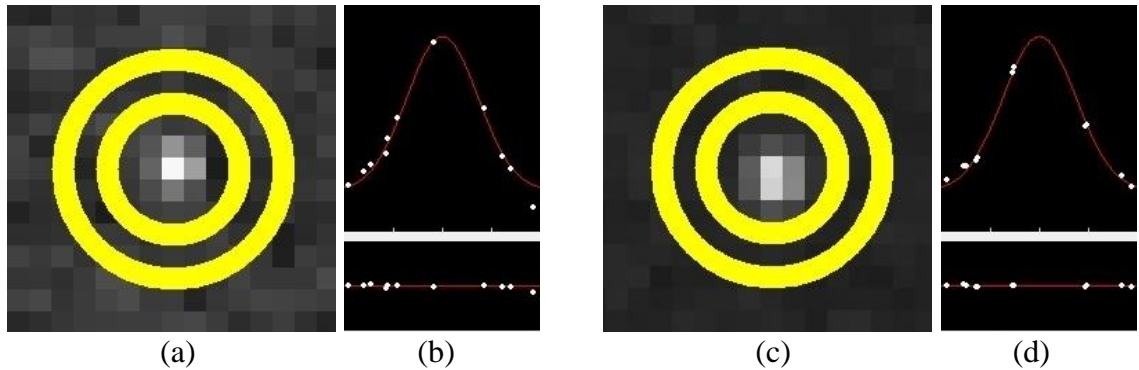

(a)  (b)  (c)  (d)

**Figure 4.** (a), (c) Enlargements from a calibration image showing the pixels displayed in MaxIm DL. The rings define the calculation regions. (b), (d) Images show the results from curve fitting in Astrometrica with the residual error in the lower portion. The centroid parameters reported by these two programs are:
(a). (2435.189, 1464.883), SNR = 13.0, Flux = 2432, FWHM = 1.432 pixels
(b). (2435.15, 1464.90), SNR = 21.3, Flux = 2398, FWHM = 3.6 arcsec
(c). (2135.177, 1394.482), SNR = 47.7, Flux = 7953, FWHM = 1.621 pixels
(d). (2135.18, 1394.48), SNR = 35.8, Flux = 6505, FWHM = 3.7 arcsec

Because the stellar images typically measured 1.5 pixels FWHM, the pixel phase effect error [Kavaldjiev and Ninkov (1998)] was less than the errors due to turbulence and centroid measurement noise. Corrections for this effect were not necessary and were not implemented. The lenslets used in the KAI-08051 sensor also helped to mitigate pixel phase errors.

*2.4 Optical Distortion Corrections*

Modern refractor telescopes are designed to produce diffraction-limited images across a wide color spectrum, but small problems arise for stars not close to the optical axis. The relationship between a star's location, measured in spherical coordinates on the sky, and its position on the flat focal plane, measured in Cartesian pixel coordinates, is not linear over a wide field of view. This causes an angular measurement error. This optical distortion can be corrected by adding a cubic translation term to the focal plane measurements. The details of how this optical distortion correction were done for this experiment has already been published [Bruns and Bruns (2017)], so the procedure is only summarized here. The equatorial coordinates from the star catalog were converted to standard coordinates using the common trigonometric formulas. The measured centroids from the images were modelled using polynomial equations in both axes, including terms up to the third power. The difference between the centroid represented by the cubic equations and the star standard coordinates is the position error for each star. The polynomial coefficients were adjusted to minimize the least-squares sum over all of the stars in the nighttime calibration images and these coefficients were applied to the images taken during the eclipse.

The stars in the outer parts of the image, where the cubic optical distortion was greatest, had gravitational deflections less than 0.5 arcsec. For best results, the distortion needed correction to nominally 0.02 arcsec. Thus, one of the critical parts of this experiment was to calibrate the optical distortion in the telescope. Optical ray traces provided by Al Nagler of Tele Vue Optics suggested a cubic distortion of -12·$10^{-16}$ rad/pixel$^3$. This gave 1.0 arcsec distortion near the right edge of the sensor at 1600 pixels. Small manufacturing or alignment errors slightly modified this coefficient, based on the measurements described next.

The optical distortion measurements were made at night. Measurements made in March 2017 in San Diego were at an average air temperature of 13 C and distortion measurements at the eclipse setup site on two evenings before the eclipse (August 18 and August 19) averaged 11 C. These were both close enough to the eclipse-day data taken near 13 C that temperature effects were not included. The procedure was to point the telescope to the same altitude and azimuth where the eclipse would occur and take ten images, each 10 s long, while the telescope was accurately tracking the stars. After a few minutes the telescope was slewed back to the starting position and ten more images were acquired. This procedure was repeated 30 times over two evenings. Since new stars drifted into the field of view after each repointing, the calibration images built up a dense, random pattern of stars that were analyzed for distortion.

The results for the August 2017 tests closely matched the results from the March 2017 tests, confirming the stability of the telescope optics. Table 2 shows the final results for all of the cubic coefficients and Figure 5 graphs the resulting distortion field. The measured distortion was about two-thirds of the value calculated from the optical ray traces. Also shown in the figure is the difference in the centroid shifts between the values calculated by Astrometrica and the values calculated by averaging the Astrometrica and the PinPoint coefficients. The stars used in the final eclipse analysis are overlaid on the contour plot, showing that for all but three stars, the difference in the distortion shift is less than 0.010 arcsec. These are small compared with the RMS fitting errors of 0.065 arcsec. The optical distortion in all of the master images was corrected by applying the cubic coefficients shown in Table 2.

Table 2. Optical Distortion coefficients. The first column indicates the coefficient powers associated with the various cubic terms. The second and fourth columns give the coefficients measured in August 2017. The third and fifth columns give the amplitude of the distortions near the edges of the sensor field of view. The $X^2Y$ and $Y^3$ terms in RA and the $X^3$ and $XY^2$ terms in Dec result in relatively small corrections.

| Coefficient description | X-axis (RA) (rad/pixel$^3$) | X-axis (RA) at 1600/1200 pixels (arcsec) | Y-axis (Dec) (rad/pixel$^3$) | Y-axis (Dec) at 1600/1200 pixels (arcsec) |
|---|---|---|---|---|
| $X^3$ | -7.96·$10^{-16}$ | -0.673 | -0.63·$10^{-16}$ | -0.053 |
| $X^2Y$ | -0.82·$10^{-16}$ | -0.052 | -3.46·$10^{-16}$ | -0.219 |
| $XY^2$ | -4.47·$10^{-16}$ | -0.213 | +0.23·$10^{-16}$ | +0.011 |
| $Y^3$ | +0.20·$10^{-16}$ | +0.007 | -9.24·$10^{-16}$ | -0.329 |

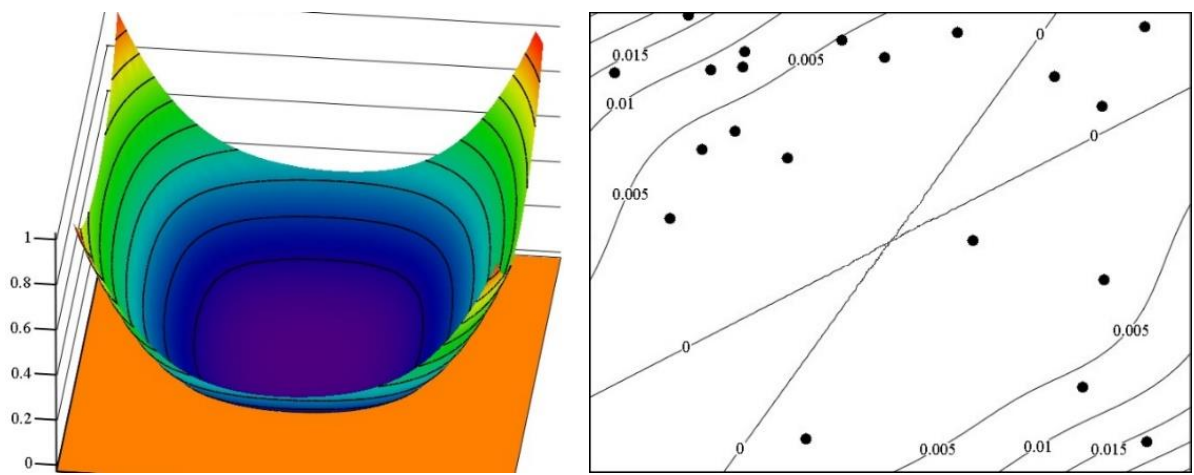

**Figure 5.** (Left) The distortion across the sensor is plotted as contours relative to the image center. The vertical scale is in arcseconds and the surface contours are marked in 0.1 arcsec intervals. Near the corners, the distortion amplitude reaches 1 arcsec. The contour curves are not circular, but slightly flattened due to the small cubic cross-terms. (Right) The difference between the distortion amplitude measured using only the Astrometrica centroids and the average of the Astrometrica and PinPoint centroids is plotted on a contour diagram with intervals of 0.005 arcsec. The stars used in the final solution are indicated with the dots.

## 2.5 Eclipse Day Procedure

The weather on Casper Mountain was nearly ideal during the eclipse. A few thin clouds were seen approaching the Sun one hour before totality, but they did not affect imaging. Winds were calm. The ambient temperature during totality fell from 13.4 C down to 13.1 C as indicated in the next figure. During totality the relative humidity measured 40% ± 2%, indicating a dew point of 2.3 C. The absolute barometric pressure measured 770.1 mbar, in agreement with the Casper airport reading (corrected to sea level) of 1017.8 mbar.

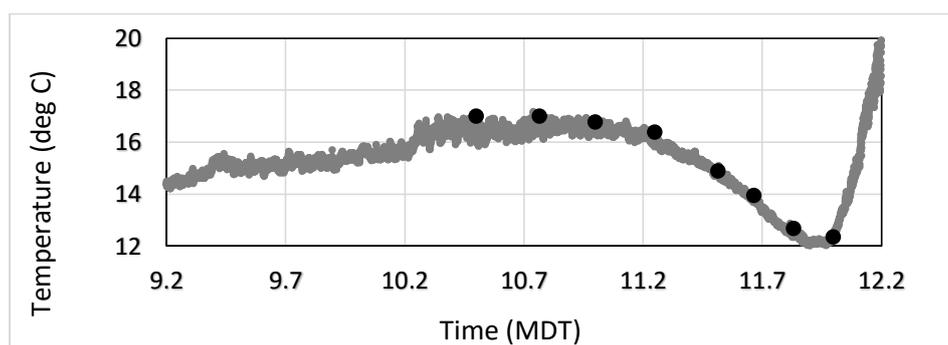

**Figure 6.** The ambient temperature was monitored about 10 m from the telescope, at 2 m above ground level. The graph shows the ambient temperature measured with two different instruments. The large dots are manually recorded from a sensor with a 1-minute response time in air. The smaller points are from an automatic recording set for two second intervals using a smaller sensor that has a 1 second response time in air. Mid-totality occurred at 11.73 MDT, when the temperature was 13.3 C. At about noon, the sensor was partially exposed to direct sunlight.

The telescope was polar aligned to the true (non-refracted) pole to within 4 arcmin using TheSkyX's T-point software routines. The camera was rotated so that the rows were parallel to the RA axis within 0.4°. This simplified the optical distortion measurements noted in the previous section.

In the descriptions here, as well as in the following sections, the field surrounding the Sun (but offset 0.3° in declination north of it) is defined as the ECLIPSE field. It was chosen to maximize the number of measureable stars. The star field at a location 7.4° west of the Sun is called the RIGHT calibration field. It has nearly the same altitude and azimuth as the ECLIPSE field so atmospheric refraction corrections are minimized. The star field 7.4° east of the Sun is defined as the LEFT calibration field. It was chosen to be symmetrically opposite the Sun in order to minimize rotation corrections. Table 3 defines the coordinates used in these three fields and Figure 7 shows their relationship on the sky.

Table 3. Locations of the three fields used for calibrations and deflection measurements. The altitude values include refraction corrections.

| Parameter | RIGHT Field | ECLIPSE Field | LEFT Field |
|---|---|---|---|
| Center RA (degrees) | 144.048 | 150.954 | 157.858 |
| Center Dec (degrees) | 9.194 | 12.193 | 15.097 |
| Center Alt (degrees) | 54.22 | 54.38 | 53.38 |
| Center Azimuth (degrees) | 155.43 | 142.71 | 130.41 |

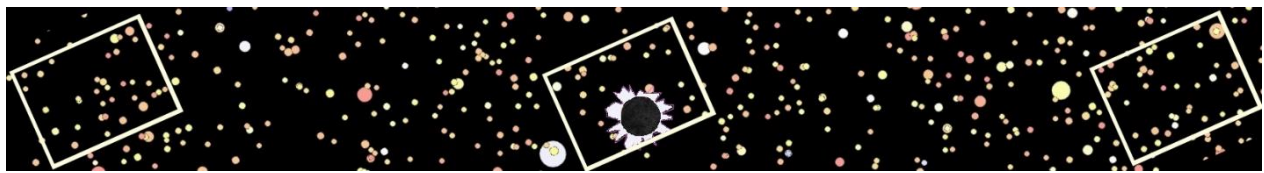

Figure 7. The LEFT, ECLIPSE, and RIGHT star fields are shown with constant altitude along the horizontal axis. Stars are colored according to their spectral class. Stars down to magnitude 10.5 are shown with brighter stars indicated with larger symbols. Image is modified from Project Pluto's Guide 9 software.

To simplify the data acquisition, the telescope pointing and the camera exposures were automated and triggered from the laptop clock synchronized to UT within one second. The final focus on Mag 3.5 Subra (6° from the sun) was performed manually about five minutes before totality. The focus mechanism was then locked and the automated script took over the data acquisition.

Starting at one minute before totality with the telescope pointing toward the RIGHT calibration field, 16 exposures were made, each 2.00 s long. An analysis based on a 2006 calibrated eclipse image [Viladrich (2016)] predicted this would be the best exposure and would allow good star measurements. This resulted in slightly overexposed images at the start but well-exposed images as totality approached. By starting before totality, this technique provided two additional calibration images.

At five seconds into totality, a 0.10 s long autoexposure image was recorded. The mean background level was calculated over a small pre-selected region. Delaying this test exposure by five seconds after the start of totality assured that the sky was completely dark and would stay relatively constant. Based on the autoexposure image, the script calculated the exposure needed to make the background level 10,000 ADU counts. The result was an exposure time of 2.28 s for the remaining 25 s spent on this calibration field, which was very close to the predicted value of 2 s. The resulting peak pixel value of every star used in the final analysis was less than 42,000 ADU counts, well within the CCD's linear range. Six additional images were acquired during totality. With the two useable frames acquired before totality, the total integrated exposure was 17.68 s.

At 32 s into totality the script commanded the telescope to move to the ECLIPSE field. While the telescope was settling, a second autoexposure image was taken and the mean background level near star SAO 98893 (Mag 9.1) was calculated. This star was chosen as the closest bright star likely to be measureable, so the background level was set here to 20,000 ADU counts. The calculated exposure was 0.62 s. The sky brightness was not expected to change during the middle of totality so the exposure was fixed at 0.62 s for the next 35 s. This technique worked well; no stars in the field had a peak pixel value exceeding 24,000 ADU counts. The total integrated exposure for the 17 frames in this series was 10.54 s.

In the middle of totality, 18 s were devoted to taking exposures 0.15 times as long as the first ECLIPSE field images in an attempt to record two bright stars located only 0.5 solar radii from the limb. This multiplier was based on the same calibrated 2006 eclipse image from Viladrich. The resulting exposure time ended up at 0.09 s, giving a total integrated exposure for these 11 frames of only 0.99 s. However, the deflections for these stars is so large compared to every other measured star that the time spent to acquire these images was predicted to be worthwhile. Results of these two stars are included in the final deflection calculation using an alignment technique described below.

The next 35 s repeated the first ECLIPSE field measurements with 0.62 s exposures, adding another 17 frames with 10.54 s of integrated exposure. The script then commanded the telescope to point to the LEFT calibration field. While the telescope was settling, another autoexposure image was recorded. The calculated exposure increased to 3.15 s. This longer exposure time resulted from the slightly darker sky measured closer to mid-totality. During the next 27 s, six useful images were saved. The script continued taking images of this field for 60 s past the end of totality using the same exposure time, but only the first one was not too bright. A total exposure time of 22.05 s was obtained for the LEFT calibration field.

Dark frames and sky flats were taken immediately following totality. These last images are used to correct pixel-to-pixel intensity variations inherent in the sensor as well as from any dust artifacts and optical vignetting. This improves sensor readout linearity which results in more accurate centroid measurements. Since the sky was expected to brighten rapidly, new autoexposure images were made after every three frames. A total of 21 useful flats were recorded in the next 65 seconds with exposures starting at 0.09 s and ending with 0.03 s. To avoid star artifacts, the telescope was automatically moved 54 arcsec in RA between each exposure. These images were averaged and then the mean pixel intensity was rescaled to 1.00, resulting in a calibration image used to correct the pixel responses across the image. The RMS of this calibration image was only 0.01, indicating a clean sensor with minimal optical vignetting.

The script paused and signaled to install the lens cap and LED source for the optical axis measurements (see Bruns and Bruns (2017)), then re-pointing the telescope to the three orientations used during totality. The LED was then turned off and 100 dark frames were recorded. This ended the script 7.5 minutes after the end of totality.

*2.6 Data Analysis*

The processing for all of the RIGHT and LEFT calibration images were similar. The brightest 17 stars were first manually located in each image using MaxIm DL. In addition to a small telescope tracking error, each star wandered slightly between exposures because of atmos-pheric turbulence. Thus, an average centroid was calculated over all of those stars. This provided a means to determine the tracking error that was insensitive to turbulence. Each frame was then shifted in MaxIm DL using bicubic pixel interpolation to align the images to sub-pixel accuracy, forming a master image. This master image was next processed in Astrometrica to find every star with an SNR greater than 20. The same stars were identified in MaxIm DL. Some stars were then eliminated, based on poor Astrometrica fits or close neighbor stars, leaving about 60 stars in each master image. The optical distortion was corrected by using the August 2017 cubic coefficients (shown in Table 2) in the polynomial equations, leaving only linear and quadratic plate scale terms to fit these calibration images.

The ECLIPSE images required a preliminary step before the centroids could be found because the corona caused a steep gradient in the image brightness that seemed to skew some centroid positions up to 0.1 arcsec. The 34 and 11 ECLIPSE image series were averaged without translations, then a 10-pixel wide Gaussian blur was applied. This effectively hid the stars but preserved the local coronal shape and brightness. This blurred image was then subtracted from all of the individual ECLIPSE images, making the centroids easily found by using Astrometrica and MaxIm DL. For the 0.62 s exposures, the brightest 17 stars were used to determine the average centroids. Only 10 stars were bright enough to be used in the 0.09 s exposures. After translation and averaging, the master frames were re-processed with the programs to get a preliminary list of centroids. Several stars were eliminated because of very poor fits or nearby stars that skewed the centroids. One bright star was located right on the edge of a bright coronal streamer, so its centroid was not used. Figure 8 shows the master and corona-subtracted images.

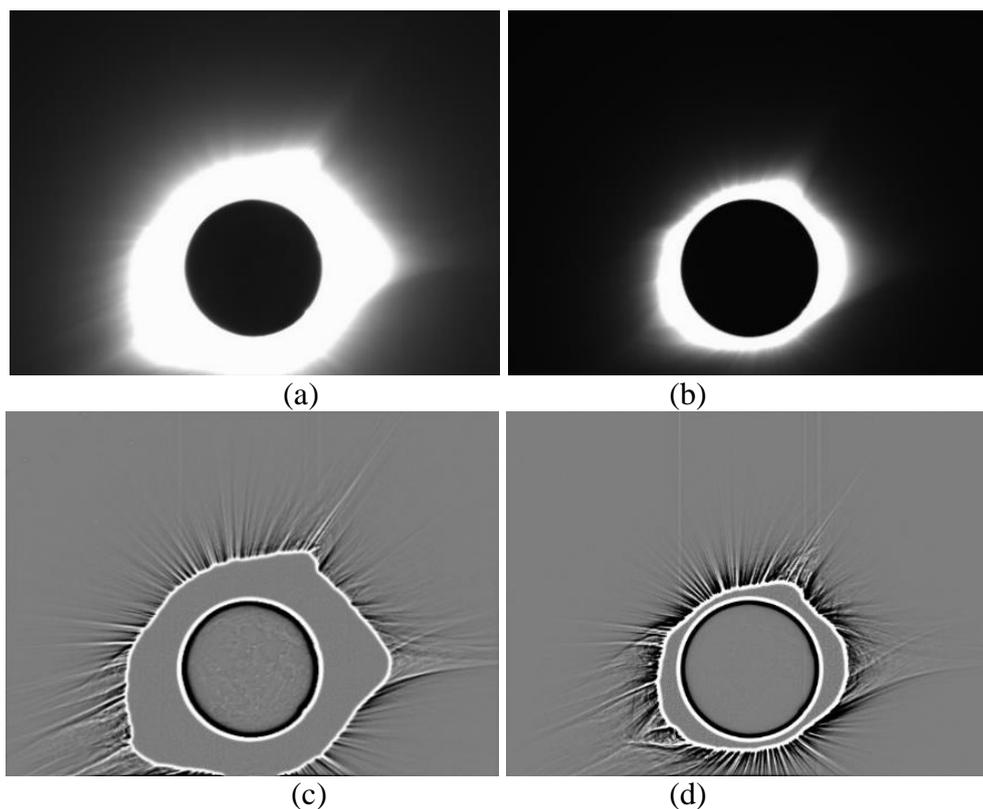

(a)  (b)

(c)  (d)

**Figure 8.** (a) ECLIPSE image with 0.62 s exposure. The Sun is slightly offset toward the bottom so that several bright stars along the upper edge were captured. Because every star centroid is less than two pixels diameter, individual stars are not seen in these images at this resolution. (b) ECLIPSE image with 0.09 s exposure. (c) and (d) By subtracting a copy of the blurred corona from the ECLIPSE image and stretching the contrast, star SNRs are enhanced and centroids easily measured.

The next step was to identify the stars with the UCAC5 catalog [Zacharias et. al. (2017)]. That catalog includes proper motion, with nominal 0.002 arcsec position errors. This accuracy was achieved by incorporating the 2016 Gaia data release [Lindegren et. al. (2016)]. Proper motion and atmospheric refraction corrections to that subset were applied using the FORTRAN version of USNO's NOVAS program, slightly modified to improve accuracy by incorporating Stone's refraction formulas [Stone (1966)]. The NOVAS program includes theoretical gravitational deflections, but this was subtracted so that deflection could be used as a variable when minimizing the difference between the centroids and the catalog positions. This means the deflections are assumed to follow the correct hyperbolic law.

The third step was to determine the plate scale and the other second-order plate constants from the two calibration fields. About 50 stars in both the RIGHT and LEFT images were found in the UCAC5 catalog. Since the image centers are 28 solar radii away from the Sun, the differential gravitational deflection is only 0.011 arcsec across the images. If the deflection constant measured in this experiment ended up within 10% of the theoretical value, then the differential error would be only 0.001 arcsec and could be ignored. Because the final value was an even closer match to the theoretical value, this differential gravitational correction was neglected.

All of the plate constants were averaged over the RIGHT and LEFT calibration fields because the ECLIPSE field was midway between them. Those mean values are shown in Table 4. These plate constants were then used in polynomials used in the ECLIPSE field images. Differences in RA and Dec between the centroids and the adjusted catalog positions give the initial deflection errors. Small corrections due to the spherical coordinate system and projection of the errors along the vector pointed away from the Sun gave the final deflection error for each star. The sum of the squares of these errors was minimized by trial and error, adjusting only the deflection coefficient and simple RA and Dec offsets. The Astrometrica and MaxIm DL series were optimized separately.

**Table 4.** Mean plate constants determined from the calibration fields for Astrometrica (AST) and MaxIm DL (MDL). The first rows are the linear terms that give plate scale while the final rows are the quadratic terms that correspond to image plane tilt and rotation.

| Coefficient description | X-axis (RA) (rad/pixel) (AST) | X-axis (RA) (rad/ pixel) (MDL) | Y-axis (Dec) (rad/pixel) (AST) | Y-axis (Dec) (rad/pixel) (MDL) |
|---|---|---|---|---|
| **X (plate scale)** | $-1.011697 \cdot 10^{-5}$ | $-1.011706 \cdot 10^{-5}$ | $-7.412 \cdot 10^{-8}$ | $-7.415 \cdot 10^{-8}$ |
| **Y (plate scale)** | $7.462 \cdot 10^{-8}$ | $7.464 \cdot 10^{-8}$ | $-1.011617 \cdot 10^{-5}$ | $-1.011629 \cdot 10^{-5}$ |
| | **(rad/pixel$^2$) (AST)** | **(rad/ pixel$^2$) (MDL)** | **(rad/pixel$^2$) (AST)** | **(rad/pixel$^2$) (MDL)** |
| **X$^2$ (plate tilt)** | $0.259 \cdot 10^{-13}$ | $-0.140 \cdot 10^{-13}$ | $4.374 \cdot 10^{-13}$ | $3.154 \cdot 10^{-13}$ |
| **XY (rotation)** | $1.747 \cdot 10^{-13}$ | $1.715 \cdot 10^{-13}$ | $-0.426 \cdot 10^{-13}$ | $0.335 \cdot 10^{-13}$ |
| **Y$^2$ (plate tilt)** | $0.482 \cdot 10^{-13}$ | $1.134 \cdot 10^{-13}$ | $4.058 \cdot 10^{-13}$ | $2.880 \cdot 10^{-13}$ |

## 3. Results

The experiment was successfully executed as planned with no equipment failures. The 18 stars found in the 0.62 s master ECLIPSE image are summarized in Table 5 and the stars are plotted in Figure 8.

Table 5. Measured ECLIPSE star details for the MaxIm DL centroids and results (a) and the Astrometrica centroids and results (b). The distances to the Sun are in units of solar radii. The deflections and the deflection errors are the outputs from the analysis based on the final least-squares values.

5 (a)

| MDL | X (pixels) | Y (pixels) | Distance | SNR | FWHM (pixels) | Deflection (arcsec) | Deflection Error (arcsec) |
|---|---|---|---|---|---|---|---|
| 1 | 120.905 | 330.576 | 4.566 | 162 | 1.90 | 0.450 | +0.063 |
| 2 | 425.492 | 1120.897 | 3.000 | 35 | 1.77 | 0.538 | -0.051 |
| 3 | 788.888 | 646.437 | 3.058 | 19 | 1.67 | 0.585 | +0.007 |
| 4 | 527.111 | 17.981 | 4.522 | 165 | 1.71 | 0.329 | -0.062 |
| 5 | 3051.848 | 78.155 | 4.817 | 52 | 1.72 | 0.369 | +0.002 |
| 6 | 841.929 | 213.205 | 3.802 | 138 | 1.60 | 0.547 | +0.083 |
| 7 | 2815.052 | 510.911 | 3.760 | 19 | 1.60 | 0.406 | -0.064 |
| 8 | 602.019 | 746.530 | 3.168 | 73 | 1.77 | 0.584 | 0.026 |
| 9 | 651.884 | 313.647 | 3.828 | 51 | 1.76 | 0.354 | -0.107 |
| 10 | 2014.265 | 112.192 | 3.694 | 139 | 1.59 | 0.537 | +0.059 |
| 11 | 828.821 | 299.299 | 3.649 | 21 | 1.64 | 0.507 | 0.023 |
| 12 | 2553.122 | 351.481 | 3.676 | 45 | 1.74 | 0.469 | -0.012 |
| 13 | 2709.304 | 2037.461 | 2.446 | 69 | 1.76 | 0.666 | -0.056 |
| 14 | 2824.313 | 1452.939 | 2.694 | 27 | 1.63 | 0.664 | +0.009 |
| 15 | 1610.946 | 245.588 | 3.306 | 28 | 1.67 | 0.502 | -0.032 |
| 16 | 3063.587 | 2332.805 | 3.395 | 74 | 1.74 | 0.491 | -0.029 |
| 17 | 1077.274 | 793.764 | 2.433 | 26 | 1.62 | 0.786 | +0.060 |
| 18 | 1374.400 | 153.860 | 3.555 | 97 | 1.62 | 0.344 | -0.153 |

5 (b)

| AST | X (pixels) | Y (pixels) | Distance | SNR | FWHM (arcsec) | Deflection (arcsec) | Deflection Error (arcsec) |
|---|---|---|---|---|---|---|---|
| 1 | 121.87 | 330.57 | 4.566 | 78 | 4.2 | 0.408 | +0.028 |
| 2 | 425.49 | 1120.88 | 3.000 | 23 | 4.1 | 0.538 | -0.040 |
| 3 | 788.88 | 646.43 | 3.058 | 15 | 3.9 | 0.582 | +0.014 |
| 4 | 527.12 | 17.97 | 4.522 | 42 | 3.9 | 0.270 | -0.113 |
| 5 | 3051.84 | 78.15 | 4.817 | 23 | 3.9 | 0.242 | -0.118 |
| 6 | 841.92 | 213.19 | 3.802 | 49 | 3.7 | 0.552 | +0.096 |
| 7 | 2815.07 | 510.92 | 3.760 | 17 | 3.8 | 0.316 | -0.145 |
| 8 | 602.00 | 746.51 | 3.168 | 34 | 3.9 | 0.606 | +0.059 |
| 9 | 651.88 | 313.62 | 3.828 | 22 | 4.0 | 0.358 | -0.095 |
| 10 | 2014.30 | 112.18 | 3.694 | 55 | 3.5 | 0.576 | +0.106 |
| 11 | 828.82 | 299.28 | 3.649 | 15 | 3.9 | 0.511 | +0.036 |
| 12 | 2553.14 | 351.44 | 3.676 | 23 | 4.1 | 0.497 | +0.025 |
| 13 | 2709.32 | 2037.44 | 2.446 | 41 | 3.9 | 0.641 | -0.068 |
| 14 | 2824.34 | 1452.93 | 2.694 | 26 | 3.6 | 0.635 | -0.009 |
| 15 | 1610.94 | 245.59 | 3.306 | 21 | 4.0 | 0.506 | -0.019 |
| 16 | 3063.60 | 2332.84 | 3.395 | 36 | 4.1 | 0.524 | +0.013 |
| 17 | 1077.30 | 793.78 | 2.433 | 23 | 3.6 | 0.728 | +0.016 |
| 18 | 1374.42 | 153.85 | 3.555 | 39 | 3.7 | 0.362 | -0.125 |

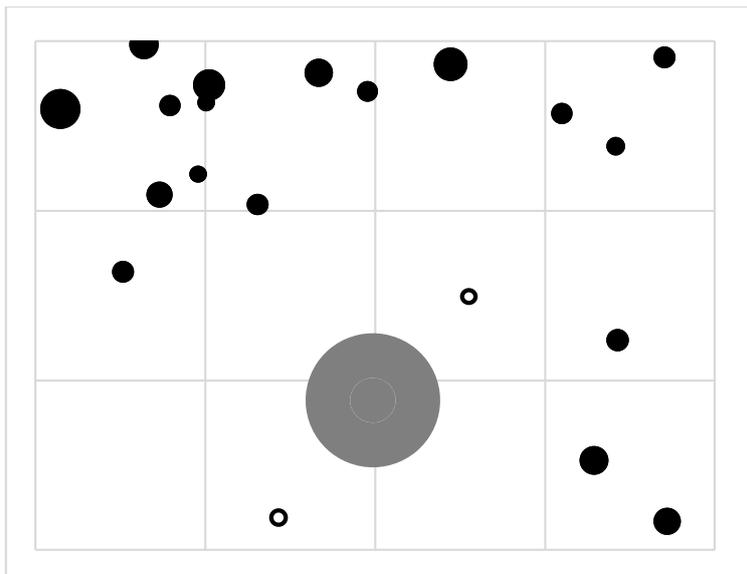

**Figure 9.** Stars used in the deflection measurements are indicated with black disks for the 0.62 s images and as an open circle for the 0.09 s images. The area of the disk indicates its SNR calculated by MaxIm DL. The large gray disk indicates the Sun's location and diameter.

Several stars have deflection errors exceeding 0.1 arcsec, but this number falls within a normal distribution. For the RMS of 0.060 arcsec for the 18 stars in the MaxIm DL series, one point is statistically probable to exceed twice the RMS, and this is the case. For the RMS of 0.073 arcsec for the 18 stars in the Astrometrica series, none of the points exceeds twice the RMS. A more detailed examination is not useful for such a small sample population.

The two close-in stars located within the corona were successfully imaged with the results shown in Table 6. Those exposures were 0.09 s and the centroid variance between the frames was 1.5 times that of the 0.62 s exposure series. Instead of including all of the stars in final analysis, only the two close-in star centroids from this series were used. Since the image center offset between the two series was random (due to small tracking errors), the average centroid of seven brightest stars in the short-exposure master image were compared to the same seven stars found in the longer-exposure master image. This produced an offset that was applied to the two close-in stars and those shifted centroids were added to the final analysis. Their final parameters are shown in the next table. The larger errors here may be due to the decreased turbulence averaging over the shorter exposures as well as the smaller SNRs caused by the bright background corona.

The deflection coefficients L that minimized the RMS of the errors for all 20 stars were 1.7338 arcsec and 1.7658 arcsec for the Astrometrica and MaxIm DL centroids, respectively. These values were averaged to obtain the final reported value of L = 1.7520 arcsec by using the mean RMS of the centroid errors (0.086 arcsec and 0.065 arcsec, respectively) as a weighting function. This is, by a wide margin, the best result for the deflection coefficient L ever obtained during an eclipse.

One of the key features that led to this remarkable result was the use of plate scale measurements based on images taken on both sides of the Sun during totality. Had this not been done in this experiment, the resulting value for the deflection coefficient would have been different by about 1%. This technique is an improvement over the historic expeditions that attempted calibration images on only one side of the sun.

**Table 6.** Measured details for the MaxIm DL centroids and results for the two close-in stars (a) and the Astrometrica centroids and results (b). The distances to the Sun are in units of solar radii. The deflections and the deflection errors are the outputs from the analysis based on the final least-squares values.

6 (a)

| MDL | X (pixels) | Y (pixels) | Distance | SNR | FWHM (pixels) | Deflection (arcsec) | Deflection Error (arcsec) |
|---|---|---|---|---|---|---|---|
| 19 | 2102.271 | 1241.101 | 1.513 | 11 | 1.44 | 1.285 | +0.118 |
| 20 | 1178.726 | 2313.957 | 1.603 | 12 | 1.53 | 1.081 | -0.020 |

6 (b)

| AST | X (pixels) | Y (pixels) | Distance | SNR | FWHM (arcsec) | Deflection (arcsec) | Deflection Error (arcsec) |
|---|---|---|---|---|---|---|---|
| 19 | 2102.28 | 1241.07 | 1.513 | 14 | 3.5 | 1.339 | +0.193 |
| 20 | 1178.75 | 2313.99 | 1.603 | 17 | 4.0 | 1.107 | -0.064 |

Figure 9 shows the deflection results plotted as a function of their distance from the Sun. The curve is the theoretical deflection using the ideal 1.751 arcsec coefficient. All of the stars are at a distance greater than 2.4 solar radii except the two stars near 1.5 solar radii. Variations in the analysis and the underlying uncertainties are discussed in the next section.

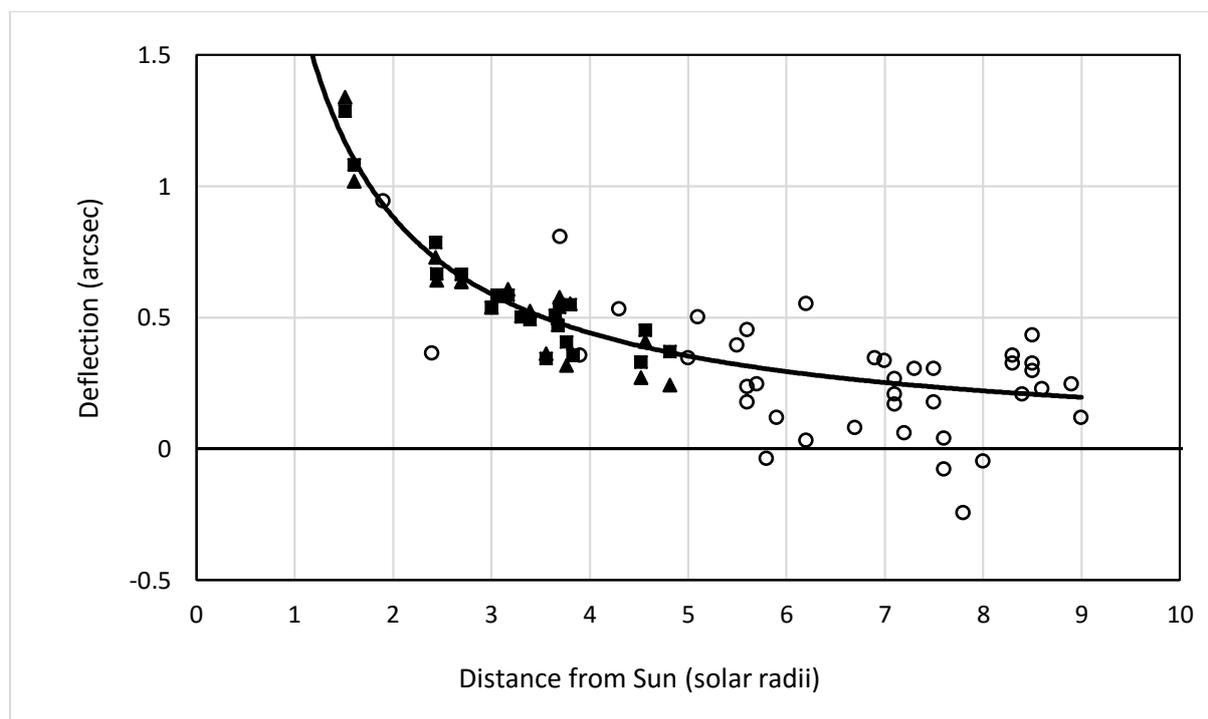

**Figure 10.** The deflection measurements for all 20 stars are plotted as a function of radial distance from the Sun. The solid curve follows the theoretical value of 1.751 arcsec. The triangles mark the Astrometrica results and the squares mark the MaxIm DL results. For comparison, the results from the 1973 experiment are shown as open circles.

## 4. Discussion

The final weighted average of L = 1.752 arcsec has an error of only 0.05%. The uncertainty in the value of L can be calculated by using the Freundlich and Ledermann equations [Freundlich and Ledermann (1944)]. They outline two methods to calculate the deflection coefficient. Their first method uses only the images taken near the Sun and determines both the plate scale and the gravitational deflections from the same stars. Their second method is the one used here, where the plate scale is independently determined from images taken far from the Sun. This technique requires calculating the uncertainty in the plate scale determined from the RIGHT and LEFT master calibration images. This was solved by performing a moment calculation.

The plate scale can be estimated by calculating the ratio of the distance between two stars separated by a known angle divided by the same distance measured in pixels. For highest accuracy, the stars should be on opposite sides of the image. Since the angular distance is based on the UCAC5 star catalog, it is much more accurate than the pixel separation. The uncertainty in the centroid position divided by the distance between the centroids gives the uncertainty in the plate scale for one star pair. Since 96 stars were used in the plate scale analysis and some of these were close to the image center, a more detailed calculation was needed to determine the final plate scale uncertainty. An analysis shows that the uncertainty in the plate scale can be expressed as a type of moment calculation, where each star has its own plate scale and centroid error. After a little algebra, the RMS error multiplied by the square root of the number of stars, divided by the sum of their distances from the image center, gives the aggregate plate scale uncertainty. For a dense, uniform array of stars, this is equal to twice the mean centroid error divided by the widest separation. For this experiment, the calculated factor is closer to 1.5, based on the distribution of stars in the images.

The RIGHT and LEFT master images have a mean RMS centroid error of 0.077 arcsec, averaged over the Astrometrica and MaxIm DL series. Using the formula in the previous paragraph leads to a relative plate scale uncertainty of 1.00000334. When scaled to the uncertainty at one solar radius (948.311 arcsec), this gives a final uncertainty of 0.00317 arcsec. According to Freundlich and Ledermann's Equation (23), this gives an uncertainty in the value of L of 1.23%. This is the first time that this separate plate scale determination has ever succeeded.

The major part of the total uncertainty in L, however, comes from the RMS fit of the stars in the ECLIPSE master image. Using Freundlich and Ledermann's Equation (20) gives the error in L, scaled to one solar radius, as 0.088 arcsec, or 3.1%. Combined with the plate scale uncertainty, the final reported uncertainty is 3.4%. This is the smallest uncertainty ever reported for this kind of experiment.

This uncertainty is based on the distribution of stars as well as the number of stars, where every star is given the same weight. If the two close-in stars are omitted in this calculation, the centroid uncertainty improves by 16%. The stellar distribution component increases by a factor of 35% when these close-in stars are not included. The end result of neglecting the two close stars increases the final uncertainty to 4.1%. The deflection coefficient L for this calculation is 1.731 arcsec, still only 1% from the theoretical value and well within the uncertainty.

An alternate method to calculate the deflections uses the same technique used by all of the previous eclipse deflection experiments. This allows an interesting experimental comparison to be made, although the equipment used in the present experiment is, of course, substantially different. This technique uses only the ECLIPSE images, calculating both the plate scale and the deflections from only those 20 stars. The reason this works is because the plate scale moves the centroids linearly from the Sun while the gravitational coefficient causes the centroids to decrease hyperbolically from the Sun. According to the Freundlich and Ledermann's

Equation (12), the uncertainty is increased to about 4% when this technique is used. Based on this analysis, the weighted coefficient is calculated to be L = 1.86 arcsec, in error from the theoretical value by 6%. If the two close-in stars are not included, the weighted coefficient is calculated to be L = 1.711 arcsec with an uncertainty of 8%. These numbers show the value of using the RIGHT and LEFT calibration fields to determine the plate scale.

Another small source of error is related to how the centroids of the two stars close to the Sun were added to the data. It turns out that since these two stars are nearly opposite the Sun, much of the effect of a small translation error is cancelled. Based on the 0.065 arcsec RMS error of the seven stars used for the alignment, using only one of the two close-in stars would contribute a deflection error of about 6%. Combining the nearly opposite deflections of the two close-in stars gives an error only 0.3%, which is a factor of 17 reduction due to their symmetrical placement. Since the deflections of these two stars must be added to the other 18 stars, the final deflection coefficient is affected less than 0.1%. This fortuitous alignment of two bright stars less than two solar radii from the Sun won't be repeated over land until the Hawaiian eclipse of 2254. The eclipses of 1919, 1954, and 1973 failed to image similarly close pairs.

The data can also be plotted using the formula suggested by Danjon [Danjon (1932)]. The measured deflections divided by their distance from the Sun are plotted along the horizontal axis. The vertical axis is 1.751 arcsec divided by the squares of the star's distances from the Sun. The slope of the best fit line gives the ratio of the measured deflection constant to the theoretical 1.751 arcsec value. This type of analysis automatically weights the stars closer to the Sun, so some researchers have objected to its use [von Klüber (1960)]; it is shown here just for completeness. Figure 10 shows the curve for the mean deflections (averaged over the Astrometrica and MaxIm DL analyses) of the 20 measured stars. The slope of this line is 1.031, resulting in a deflection value of 1.805 arcsec. The intercept at the vertical axis was constrained to zero, forcing the plot to use the accurately measured plate scale.

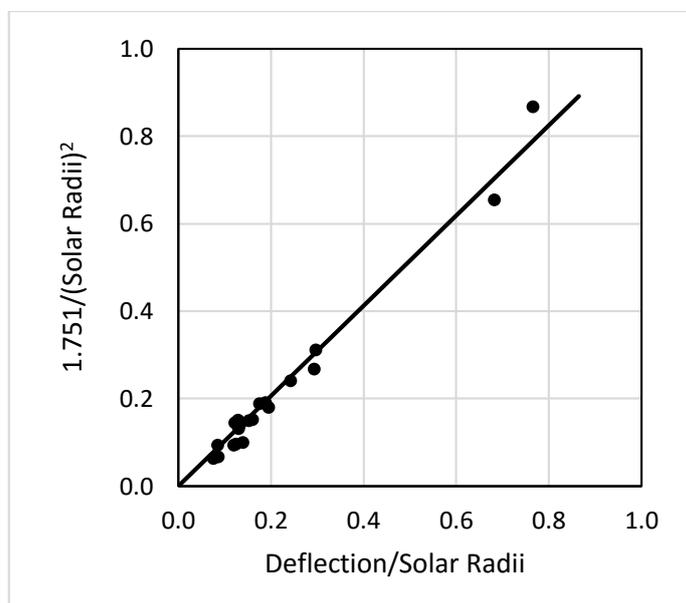

**Figure 11.** The mean of the Astrometrica and the MaxIm DL results are plotted in a Danjon graph. The slope of the line is 1.031, which gives a deflection coefficient of 1.805 arcsec. This analysis weights the stars with the largest deflections. All distances are measured in units of solar radii.

Based on the images acquired during this experiment, practical suggestions for similar attempts at future eclipses can be offered. To see if a CMOS monochrome camera would be viable, the 0.62 s ECLIPSE data series was re-processed after resampling the 16-bit FITS files down to 12 bits. The mean RMS difference in the centroids calculated by the 16-bit and simulated 12-bit images averaged only 0.03 arcsec and the final uncertainty in L was not affected. This shows that a 16-bit camera will not be required, opening the choice to other sensors with higher frame rates. Pixel full-well depth and sensor dynamic range should be considered, but the most critical item is the camera pixel diameter. A plate scale close to 2 arcsec per pixel gives a good compromise between high SNR and ability to measure the centroid to high precision. Larger telescope fields of view will require much better cubic distortion corrections, even as the gravitational deflections shrink further from the Sun. The use of an auto-exposure script successfully eliminated over- and under-exposed frames, so that should be implemented. A good polar alignment with a smooth tracking mount will be required. This will allow taking plate scale calibration images on both sides of the Sun. For the color filter, an even narrower band or deeper red filter might be chosen to further increase the exposure times, reducing the effects of turbulence. Finally, the use of a dedicated image series to measure the optical distortion coefficients should be continued, especially for larger fields of view.

Assuming that these suggestions are followed, it is possible to estimate the resulting uncertainties. A review of the background stars for the next few eclipses until 2024 show that there are typically 15 measurable stars if the same telescope and camera were used. The eclipse of 2027 offers 30 measureable stars. None of these, however, include a close-in, symmetric pair. This 2017 experiment was operated at a high altitude site with good seeing. The gain offered by an improved camera frame rate might compensate for the increased turbulence from a sea-level site. Longer eclipses will also reduce turbulence effects. Since these improvement all work as the square root of the number of stars or integrated exposure, the best one can hope for in the foreseeable future is a factor of two improvement, reducing the uncertainty to the range of 2 %. The astronomer's experience, however, may be greatly improved by the fact that this 2017 experiment showed the setup and data analysis are not too difficult compared to the previous risks and difficulties encountered by large expeditions.

## 5. Conclusions

This experiment had been carefully planned, analyzed, and tested for about 20 months prior to the eclipse date, so it has been a great satisfaction that the final data analysis was so successful. While there is no new science resulting from this experiment, the hopes of the 20[th] century astronomers have been realized.

Two different techniques were used to analyze the data. Imaging two calibration fields during totality was the best method to determine the critical plate scales. This is the first time the plate scales were successfully determined in this manner.

There are three longer and easily accessible total eclipses occurring over land in the next ten years, and this paper outlines the equipment and techniques that might be used. While technical changes in commercially available equipment during this period are anticipated, there is little reason to choose a different telescope. The Tele Vue Optics NP101is telescope mounted on the Software Bisque Paramount performed perfectly, are well matched, and are easily transportable. The best camera for future eclipse experiments is still to be determined, but the capabilities of Finger Lakes Instrumentation ML8051 was proven in this experiment. Future attempts might be improved by using the results of this experiment to help predict the best exposures. Repeating this experiment to an even higher precision might be of interest to student astronomers, especially as a learning experience in planning and executing remote field expeditions.

## Acknowledgements

The author thanks the vendors who graciously loaned the equipment necessary for this experiment: Al Nagler, Tele Vue Optics, Inc., for loan of the NP101is telescope, Gregory Terrance, Finger Lakes Instrumentation LLC, for loan of the ML8051 CCD camera, and Stephen Bisque, Software Bisque, Inc., for loan of the MyT Paramount tripod.

The author also thanks for help with the star catalog and astrometry calculations: George Kaplan and John Bangert, both formerly of USNO, and Norbert Zacharias, USNO. Thanks are also due to Greg Kinne, who assisted in the scripting software. Finally, thanks are due to Steve Lang, who built and installed the commemorative tripod base and to Gary Hazen and his staff at the Allen H. Stewart Lions Camp in Casper for providing space, utilities, and wonderful accommodations during a very hectic week for much of Wyoming.

This research has made use of FORTRAN version of NOVAS, the Naval Observatory Vector Astrometry Software package. NOVAS, the Naval Observatory Vector Astrometry Software, can be downloaded from the USNO website aa.usno.navy.mil/software/novas/novas_info.php.

This research has made use of the VizieR catalogue access tool, CDS, Strasbourg, France. The original description of the VizieR service was published in *A&AS* **143**, 23.

## References

Beckers, J.M., "On the relation between scintillation and seeing observations of extended objects," *Solar Physics* **145**, 399-402 (1993).

Brune, R.A. Jr. and the Texas Mauritanian Eclipse Team, "Gravitational deflection of light: solar eclipse of 30 June 1973 I. Description of procedures and final results," *A.J.* **81**, 452-454 (1976).

Bruns, D.G., "Using Arago's Spot to Monitor Optical Axis Shift in a Nagler-Petzval Refractor," *Applied Optics* **56**, 2074-2077 (2017).

Bruns, D.G. and C. T. Bruns, "Astrometric distortion calibration of a portable refractor," *Applied Optics* **56** (22), pp. 6288-6292 (2017).

Danjon, "Sur le déplacement apparent des étoiles au voisinage du Soleil éclipsé," *C.R.Acad. Sci.* **194**, 252 (1932).

Dyson F. W., Eddington A. S., and Davidson C., **"**A Determination of the Deflection of Light by the Sun's Gravitational Field, from Observations Made at the Total Eclipse of May 29, 1919", *Phil. Trans. R. Soc. A* **220**, 291–333, (1920).

Freundlich, E.F. and W. Ledermann, "The problem of an accurate determination of the relativistic light deflection," *M.N.R.A.S* **104**, 40-47 (1944).

Jones, B.F. "Gravitational deflection of light: solar eclipse of 30 June 1973 II. Plate reductions," *A.J.* **81**, 455-463 (1976).

Jorden, P. R., J.-M. Deltorn, and A. P. Oates, "The non-uniformity of CCDs and the effects of spatial undersampling," *S.P.I.E. Proceedings* **2198**, pp. 836-850 (1994)

Kavaldjiev, D. and Z. Ninkov, "Subpixel sensitivity map for a charge-coupled device sensor," *Opt. Eng.* **37** (3) pp. 948-954 (1998).

Lindegren, L. "Atmospheric limitations of narrow-field optical astrometry," *Astron. Astrophys.* **89**, pp 41-47 (1980).

Lindegren, L. and Gaia Collaboration, "Gaia Data Release 1," *Astronomy & Astrophysics* **595**, A4 (2016).

Seykora, E.J. "Solar scintillation and the monitoring of solar seeing," *Solar Physics* **145**, 389-397 (1993).

Stone, R.C. "An Accurate Method for Computing Atmospheric Refraction," *Pub. Ast. Soc. Pac.* **108**, pp. 1051-1058 (1996).

Viladrich, C., personal communication (2016).

von Klüber, H., "The Determination of Einstein's Light-Deflection in the Gravitational Field of the Sun," Vistas in Astronomy **3**, 47-77 (1960).

Will, C.M. "The 1919 measurement of the deflection of light," *Class. Quantum Grav.* **32**, (2015).

Zacharias, N. "Measuring the atmospheric influence on differential astrometry: a simple method applied to wide-field CCD frames," *P.A.S.P* **108**, pp. 1135-1138 (1996).

Zacharias, N., C. Finch, and J. Frouard, "UCAC5: New proper motions using *Gaia* DR1," *AJ* **153** 166 (2017).